\documentclass[journal]{IEEEtran}
\usepackage{bm} % bold math
\usepackage{amsmath}
\usepackage[dvips]{graphicx}

\begin{document}
\title{
Effective Resistance Mismatch 
and Magnetoresistance of a CPP-GMR system with Current-Confined-Paths
}

\author{Jun~Sato,
Katsuyoshi~Matsushita and
Hiroshi~Imamura$^{\ast}$\\
Nanotechnology Research Institute (NRI),\\
Advanced Industrial Science and Technology (AIST),\\
AIST Tsukuba Central 2, Tsukuba, Ibaraki 305-8568, Japan.
 \thanks{$^{\ast}$Corresponding author. Email address: h-imamura@aist.go.jp}}

\markboth{Journal of \LaTeX\ Class Files,~Vol.~6, No.~1, January~2007}%
{Shell \MakeLowercase{\textit{et al.}}: Bare Demo of IEEEtran.cls for Journals}

\maketitle

\begin{abstract}
%\boldmath
We theoretically study the magnetoresistance of a CPP-GMR system with
current confined paths (CCP) in the framework of Valet-Fert theory.
The continuity equations for charge and spin currents are numerically
solved with the three-dimensional CCP geometry by use of finite
element method.  It is confirmed that the MR ratio is enhanced by the 
CCP structure, which is consistent with the experimental results.
Moreover, we find that there exists a certain contact width which
maximize the MR ratio.  We show that the contact width which maximize
the MR ratio is well described by the effective resistance matching.
\end{abstract}


\IEEEpeerreviewmaketitle

%============================================================
% Introduction
%============================================================
\IEEEPARstart{C}{urrent}-perpendicular-to-plane giant
magnetoresistance (CPP-GMR) has attracted much attention for its
potential application as a read sensor for high-density magnetic
recording \cite{pratt91}, \cite{reilly99}. In order to realize a high-density magnetic
recording, we need MR devices with high MR ratio and low resistance
area product (RA). Although the RA value of a CPP-GMR system is much
smaller than that of a tunneling magnetoresistance (TMR) system, the
MR ratio of a conventional CPP-GMR system still remains a small value
of a few $\%$.  Much effort has been devoted to increasing the MR
ratio of the CPP-GMR system.

Recently, Fukuzawa {\it et al.} reported that they achieved the MR ratio of
10.2 $\%$ by CPP-GMR spin-valve with a current-confined-path (CCP)
structure made of a nano-oxide-layer (NOL) with a lot of small
metallic channels \cite{fukuzawa04}, \cite{fukuzawa05}. 
They showed that the MR ratio increases
with increasing the RA value, which implies that the MR ratio is
enhanced for the system with narrow metallic channels.  The similar
enhancement of the MR ratio due to the CCP structure containing a
domain wall was also reported by Fuke {\it et al.} \cite{fuke07}.  The
enhancement of the MR ratio due to the CCP structure is theoretically
explained by one of the author \cite{imamura07}.  However, the
analysis in \cite{imamura07} is based on the assumption that the
thickness of the non-magnetic layer is zero and little attention is
paid to the effect of the nonmagnetic spacer layer on the enhancement
of the MR ratio.

In 2000, Schmidt {\it et al.} pointed out that the conductivity mismatch is
a basic obstacle for spin injection from a ferromagnetic metal into a
semiconductor \cite{schmidt00}. Spin polarization of the injected
current is strongly suppressed by the conductivity mismatch between
these materials.  For conventional CPP-GMR spin-valves without CCP
structures, the conductivity mismatch seems not to be important
because both of the ferromagnetic electrodes and the nonmagnetic
spacer layer are made of metals.
However, for the CCP-CPP-GMR system the effective resistance of the
contact region increases with decreasing the radius of the contact,
which implies that the conductivity mismatch plays an important role
in the CCP-CPP-GMR system.
Therefore, it is intriguing to ask how the
mismatch of the effective resistance affects the MR ratio of the
CCP-CPP-GMR system.

%============================================================
% In this paper
%============================================================
In this paper, we analyze the dependence of the MR ratio of the CCP-CPP-GMR
spin-valves on the contact radius and resistance area product (RA) 
in the framework of Valet-Fert theory \cite{valet93}.
The spin-dependent electro-chemical potentials are obtained by
numerically solving continuity equations for charge and spin currents
with the three-dimensional CCP geometry by use
of the finite element method \cite{ram-mohan,ichimura07}.  
We show that the CCP-CPP-GMR spin-valve can take a larger MR ratio
than that of the conventional CPP-GMR spin-valve. 
However, the MR ratio is not a monotonic function of the contact
radius but takes a maximum value at a certain value of the contact radius.
We also show that the contact radius which maximizes the MR ratio can
be explained by considering the matching of effective resistances.

%============================================================
% Valet-Fert Theory
%============================================================
\section{Valet-Fert Theory of CPP-GMR}
By starting with the Boltzmann equation, Valet and Fert
constructed the macroscopic model of the CPP-GMR\cite{valet93}, on
which our calculation is based. Before showing our results, we shall
give a brief introduction to the Valet-Fert theory of CPP-GMR.

In the macroscopic model of the CPP-GMR, it is assumed that the
conduction electrons with spins
$s=\uparrow,\downarrow$ can be characterized by the conductivity
$\sigma_{s}$, spin relaxation rate $\tau_{s}$, density of states
$N_{s}$ and diffusion constant $D_{s}$.  
Since we consider a
collinear alignment of the magnetizations we neglect the mixing of
spin-up and spin-down bands.
The current density for each spin band under an electric field $\bm{E}$
is given by 
\begin{equation}
  \bm{j}_s
  =
  \sigma_{s}\bm{E}
  -q D_{s} \bm{\nabla} \delta n_{s},
\label{eq:cd1}
\end{equation}
where $q$ is the charge of an electron and $\delta n_{s}$ represents
the accumulation of the electrons with spin $s$.
By using Einstein's relation $\sigma_s=q^2N_sD_s$, we can rewrite
\eqref{eq:cd1} as
\begin{equation}
  \bm{j}_s
  =
  -\frac{\sigma_{s}}{q}
  \left(
    -q\bm{E}
    +\frac{ \bm{\nabla} \delta n_{s}}{N_{s}}
    \right).
    \label{eq:cd2}
\end{equation}
Introducing the deviation of the spin-dependent chemical potential from the
equilibrium state $\delta\mu_{s}$, the accumulation $\delta n_{s}$ can
be expressed as $\delta n_{s}=N_{s}\delta\mu_{s}$. Since the electric
field is given by the gradient of the electrostatic potential $\phi$ as
$\bm{E}=-\bm{\nabla}\phi$, it is convenient to define the
spin-dependent electro-chemical potential as
$\mu_{s}=\delta\mu_{s} + q\phi$.  Hence the current density
\eqref{eq:cd1} can be written as 
\begin{equation}
  \bm{j}_s
  =
  -\frac{\sigma_{s}}{q}\bm{\nabla}\mu_{s}.
\label{eq:cd3}
\end{equation}

In the ferromagnetic materials, the spin-polarization of the conductivity 
is represented by the parameter $\beta$ which is defined as
$\beta=(\sigma_{\uparrow}-\sigma_{\downarrow})/(\sigma_{\uparrow}+\sigma_{\downarrow})$. 
For example, the spin-polarization parameter $\beta = 0.36 \sim 0.5$
for Co \cite{mott}

The equations that determine electro-chemical potentials are
derived by the continuity equations for charge and spin.
The continuity equation for charge is given by 
\begin{equation}
\bm{\nabla}\!\cdot\!(\bm{j}_{\uparrow}+\bm{j}_{\downarrow})=0,
\label{eq:cons1}
\end{equation}
which means that there is no source for electron charge.
The continuity equation for spin is given by
\begin{equation}
\bm{\nabla}\!\cdot\!(\bm{j}_{\uparrow}-\bm{j}_{\downarrow})
=q\left(\dfrac{N_{\uparrow}}{\tau_{\uparrow}}\mu_{\uparrow}-\dfrac{N_{\downarrow}}{\tau_{\downarrow}}\mu_{\downarrow}\right), 
\label{eq:cons2}
\end{equation}
which means that the spin relaxation plays a role of spin source.
Substituting \eqref{eq:cd3} into \eqref{eq:cons1} and \eqref{eq:cons2}
we obtain the following diffusion equations for electro-chemical
potentials

\begin{align}
\label{eq:diff1}
&\sigma_{\uparrow}\nabla^{2}\mu_{\uparrow}+\sigma_{\downarrow}\nabla^{2}\mu_{\downarrow}=0,\\
\label{eq:diff2}
&\nabla^{2}\left(\mu_{\uparrow}-\mu_{\downarrow}\right)
=\dfrac{1}{\lambda^2}(\mu_{\uparrow}-\mu_{\downarrow}),
\end{align}
where $\lambda$ is the characteristic length of the spin diffusion 
called ``spin diffusion length''.
The spin diffusion length is defined as $\lambda=\sqrt{\tau D}$,
where $\tau$ is the averaged spin relaxation time defined as
$\tau^{-1}=(\tau_{\uparrow}^{-1}+\tau_{\downarrow}^{-1})/2$ and $D$ is
the averaged diffusion constant defined as
$D^{-1}=(N_{\uparrow} D^{-1}_{\downarrow}+N_{\downarrow}
D^{-1}_{\uparrow})/(N_{\uparrow}+N_{\downarrow})$.

By solving \eqref{eq:diff1} and \eqref{eq:diff2} with proper
boundary conditions, we obtain the electrochemical potential and
therefore the total resistance of the system. The MR ratio is defined
by the total resistance 
for the parallel alignment of the magnetizations $R_{\text{P}}$, and 
that for the anti-parallel alignment $R_{\text{AP}}$ as
$\text{MR}=(R_{\text{AP}}-R_{\text{P}})/R_{\text{P}}$.

%============================================================
% Valet-Fert Theory
%============================================================
\section{CPP-GMR without CCP}
\begin{figure}
\centerline{\includegraphics[width=\columnwidth]{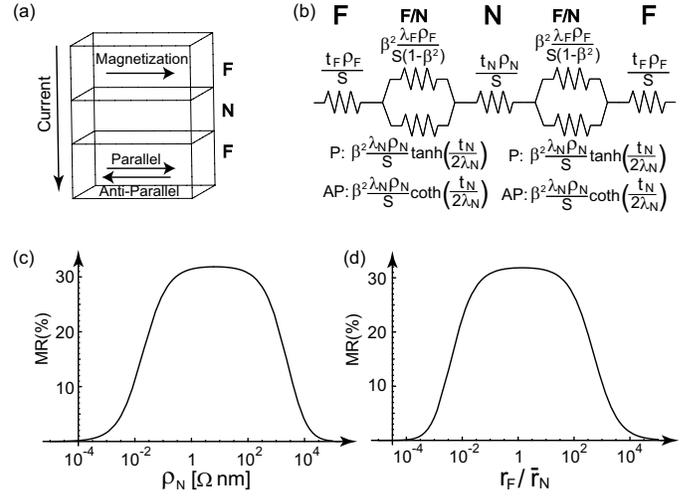}}
\caption{
  (a) A conventional CPP-GMR is schematically shown.
  The nonmagnetic (N) spacer layer is sandwiched 
  between two ferromagnetic (F) electrodes. The current is flowing
  perpendicular to the plane.  The magnetization of the top
  electrode is assumed to be fixed at a certain direction and the
  magnetization of the bottom electrode is aligned to be parallel or
  anti-parallel to that of the top electrode.
  (b) Equivalent circuit diagram representing the total resistance of
  the conventional CPP-GMR spin-valve is shown. 
  The resistance due to spin accumulation $\Delta R$ can 
  be expressed as a parallel circuit of the effective resistances. 
  (c) MR ratio is plotted against the resistivity of the non-magnetic
  layer $\rho_N$. 
  (d) MR ratio is plotted against the ratio of the effective
  resistances $r_F/\bar{r}_{N}$.
}
\label{fig1}
\end{figure}

Let us first consider the MR ratio of the conventional CPP-GMR
spin-valves shown in Fig. \ref{fig1} (a).  The nonmagnetic (N)
spacer layer is sandwiched  between two ferromagnetic (F)
electrodes. The magnetization of the top electrode is assumed to be
fixed at a certain direction and the magnetization of the bottom
electrode is aligned to be parallel or anti-parallel to that of the top
electrode.

%========================================
%  Figure 1
%========================================

We assume the translational invariance in the $x-$ and
$y-$directions and the current is flowing perpendicular to the plane,
i.e., in the $z-$direction.  Hence the equations to be solved reduce
to a couple of one-dimensional diffusion equations:
\begin{align}
\label{eq:diff1d1}
&\sigma_{\uparrow}\frac{\partial^{2}}{\partial z^{2}}\mu_{\uparrow}+\sigma_{\downarrow}\frac{\partial^{2}}{\partial z^{2}}\mu_{\downarrow}=0,\\
\label{eq:diff1d2}
&\frac{\partial^{2}}{\partial z^{2}}\left(\mu_{\uparrow}-\mu_{\downarrow}\right)
=\dfrac{1}{\lambda^2}(\mu_{\uparrow}-\mu_{\downarrow}).
\end{align}
The general solutions of \eqref{eq:diff1d1} and \eqref{eq:diff1d2} are
obtained as
\begin{align}
&\mu_{\uparrow}(z)
=c_1+c_2 z
+\frac{1}{\sigma_{\uparrow}}\left(c_3 e^{-z/\lambda}+c_4 e^{z/\lambda}\right),
\\
&\mu_{\downarrow}(z)
=c_1+c_2 z
-\frac{1}{\sigma_{\downarrow}}\left(c_3 e^{-z/\lambda}+c_4
  e^{z/\lambda}\right),
\end{align}
where $c_j$'s are arbitrary integration constants determined by the
proper boundary conditions. 

We neglect the effect of interfacial scattering for simplicity.  Then
we adopt the boundary conditions such that 
both $\mu_s$ and $\bm{j}_s$ are continuous at the interface. 
After some algebra, we can show that the total resistance 
for parallel alignment of the magnetizations $R_{\text{P}}$, and 
that for the anti-parallel alignment $R_{\text{AP}}$,
have the form
\begin{equation}
R_{\text{P,AP}}=R_0+\Delta R_{\text{P,AP}},
\end{equation}
where $R_0=\left(2t_F\rho_F+t_N\rho_N\right)/S$, $t_{F(N)}$ 
is the thickness of the F(N) layer, $S$ is the cross-section area
of the system, and $\rho_{F(N)}$ is the resistivity of the F(N) layer
defined as
\begin{equation}
\rho_{F} = \frac{1}{\sigma_{F}^{\uparrow} + \sigma_{F}^{\downarrow}}
\ \ 
\rho_{N} = \frac{1}{\sigma_{N}^{\uparrow} + \sigma_{N}^{\downarrow}}.
\end{equation} 
Here $\sigma_{F(N)}^{s}$ is the conductivity for electrons with spin $s$
in the F (N) layer.
$R_0$ is the total resistance of the system in the absence of the spin
accumulation and independent of the magnetization configuration. 
The additional resistance $\Delta R$ originates from the voltage
drop at the interfaces due to the spin accumulation, 
which has the form
\begin{equation}
\Delta R =2 \left\{r_F^{-1}+r_N^{-1}\right\}^{-1},
\label{eq:DR}
\end{equation}
where $r_F$ and $r_{N}$ are the effective resistances for F and N
layers defined as
\begin{equation}
r_F=\frac{\beta^{2}\rho_F\lambda_F}{S(1-\beta^2)},
 \ \ 
r_N=\frac{\beta^{2}\rho_N\lambda_N}{S} f_{\text{P,AP}}\left[\frac{t_N}{2\lambda_N}\right].
\end{equation}
The function $f_{\text{P,AP}}$ is given by $f_{\text{P}}(x)=\tanh(x)$
and $f_{\text{AP}}(x)=\coth(x)$. 
From \eqref{eq:DR}, $\Delta R$ is considered as 
a parallel circuit of the effective resistances $r_F$ and $r_N$ as 
shown in Fig. \ref{fig1} (b). 

In Fig. \ref{fig1} (c), 
we plot the MR ratio against the resistivity of the non-magnetic layer $\rho_N$. 
We assume that the F layer is made of CoFe and the N layer is made of Cu.
For numerical calculation, we use the following values: 
$\rho_F=160\,\Omega\,\text{nm}$, 
$t_F=45\,\text{nm}$, 
$t_N=2\,\text{nm}$, 
$\lambda_F=15\,\text{nm}$, 
$\lambda_N=500\,\text{nm}$, 
$\beta=0.7$. 
The thickness of the F-layer $t_F$ is taken to be thick enough
compared with the spin 
diffusion length $\lambda_F$ in order to eliminate the spin
accumulation at the bottom 
and the top electrodes.

One can obviously see that the MR ratio takes its maximum value at a
certain value of $\rho_N$ and vanishes in both limits of $\rho_N\to \infty$
and $\rho_N\to 0$. 
In the limit of $\rho_N\to \infty$, the effective resistance of
N-layer $r_N$ is much larger than that of F-layer $r_F$ and 
the interface resistance $\Delta R$ equals to $2\beta^2 r_F$ 
for both P and AP cases, which means that the MR ratio goes to zero. 
Conversely, in the limit of $\rho_N\to 0$, 
$r_N$ is much smaller than $r_F$ 
and the interface resistance $\Delta R$ goes to zero. 
Therefore the MR ratio vanishes also in this limit. 
% The equivalent circuit shown in Fig. \ref{fig1} tells us that
% the MR ratio is maximized if the effective resistance $r_F$ and $r_N$
% are balanced with each other $r_F\sim r_N$. 

In Fig. \ref{fig1} (d), the MR ratio is plotted against the
ratio of the effective resistance  $r_F/\bar{r}_N$, where
$\bar{r}_N=\rho_N \lambda_N$ is the effective resistance of N-layer 
for an F/N bilayer system.
As shown in Fig. \ref{fig1} (d), 
the MR ratio is maximized if the effective resistances $r_F$ and $\bar{r}_N$ are balanced 
with each other, i.e., $r_F\sim \bar{r}_N$. 
A similar idea is also applicable to the CCP-CPP-GMR system 
as will be discussed in the next section.

%========================================
% Fig. 2
%========================================
\begin{figure}[t]
  \centerline{\includegraphics[width=\columnwidth]{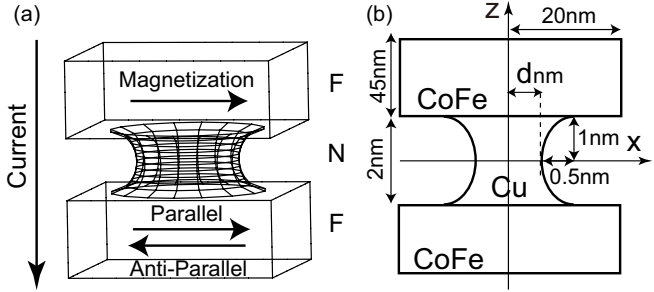}}
  \caption{
  (a) A CCP-CPP-GMR spin-valve is schematically shown.
   The nonmagnetic (N) contact region is sandwiched 
   between two ferromagnetic (F) electrodes. 
%   The current is flowing
%   perpendiculat to the plane.  
%   The magnetization of the top
%   electrode is assumed to be fixed at a cirtain direction and the
%   magnetization of the bottom electrode is alignt to be parallel or
%   anti-parallel to that of the top electrode.
  (b) The cross-sectional view of the CCP-CPP-GMR spin-valve is
  schematically shown. The curve of the contact region is modeled by
  the half ellipse,  which is tangent to the ferromagnetic layer. 
%   The length of the major axsis of the ellipse, which is the same as the
%   thickness of the contact, is taken to be 2 nm.  The length of the
%   semiminor axisis of the ellipse is 0.5 nm.
%   The F electrodes are assumed to be 40 nm x 40nm x 45 nm rectangles.
   In our calculation we keep the shape of the ellipse fixed and vary the
   size of the contact by changing the contact radius $d$.
}
  \label{fig2}
\end{figure}
%============================================================
% CCP-CPP-GMR
%============================================================
\section{CCP-CPP-GMR}
Next we move onto the CCP-CPP spin-valves, where the spacer layer
is made of an NOL with a lot of nanometer-size metallic channels.
The system we consider is schematically shown in Fig. \ref{fig2}
(a).
A narrow non-magnetic metallic contact is sandwiched 
between two ferromagnetic electrodes. 
The cross-sectional view of the system is shown in Fig. \ref{fig2} (b). 
The curve of the contact region is modeled by the half ellipse, 
which is tangent to the ferromagnetic layer. 
The length of the major axis of the ellipse, which is the same as the
thickness of the contact, is taken to be 2 nm.  The length of the
semi-minor axis of the ellipse is 0.5 nm.
The F electrodes are assumed to be 40 nm x 40nm x 45 nm rectangles.

In our calculation we keep the shape of the ellipse fixed and vary the
size of the contact by changing the contact radius $d$.
The electro-chemical potentials are determined by numerically solving
\eqref{eq:diff1} and \eqref{eq:diff2} by use of the finite element method 
\cite{ram-mohan}, \cite{ichimura07}.
The system is divided into hexahedral elements and
the total number of the elements is of the order of 10$^5$.
We assume that the F electrodes and N contact are made of CoFe and
Cu, respectively.  We use the same material parameters we used in the
previous section for the conventional CPP-GMR spin-valve system.
% $\beta=0.7$
% $\rho_{\rm CoFe}=160\,\Omega\,\text{nm}$, 
% $\lambda_{\rm CoFe}=15\,\text{nm}$
% and $\lambda_{\rm Cu}=500\,\text{nm}$.
The resistivity of Cu is assumed to be 
$\rho_{\rm Cu}=$17, 50, 170 and 300 $\Omega\,\text{nm}$.

\begin{figure}
\centerline{\includegraphics[width=0.8\columnwidth]{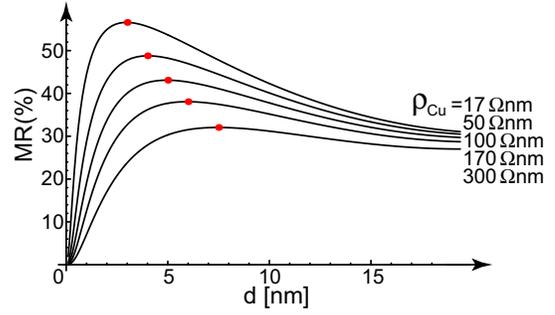}}
\caption{
  MR ratios of CCP-CPP-GMR spin-valves are plotted against the contact
  radius $d$.  
  The resistivity of Cu in the contact is varied from 17 $\Omega$nm to
  300 $\Omega$nm. 
  The MR ratio takes a maximum value at a certain value of the contact
  radius which is indicated by a circle.
}
\label{fig3}
\end{figure}

The obtained MR ratios of the CCP-CPP-GMR system are plotted against
the contact radius $d$ in Fig. \ref{fig3}. 
One can immediately see that the MR ratio increases with decreasing
the resistivity of Cu,  $\rho_{\text{Cu}}$. 
It has been considered so far that the MR ratio shows monotone increasing 
with decreasing the contact radius both theoretically \cite{imamura07}
and experimentally \cite{fukuzawa05}. 
We note that there is a contact radius that maximize the MR ratio, which is
indicated by a circle in Fig. \ref{fig3},  for
given values of $\rho_{\text{Cu}}$ and the MR ratio vanishes in
the limit of $d\to 0$.
Fig. \ref{fig3} also shows that the value of the contact radius that
maximize the MR ratio decreases with decreasing the  resistivity of Cu.
%========================================
% Fig. 3
%========================================

Strictly speaking, the Boltzmann approach loses its validity in the limit $d \to 0$, 
especially in the region $d < 1 \text{nm}$. 
In Ref. \cite{APL91}, the electron transport across a narrow metallic channel 
with the width of a single atom is studied based on the Landauer formalism, 
where it is concluded that MR ratio is diminished by the quantum interference. 
This does not contradict our results that MR goes to zero in the limit $d \to 0$. 

\begin{figure}
\centerline{\includegraphics[width=\columnwidth]{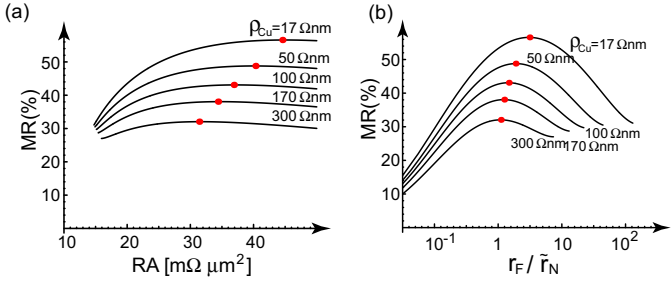}}
\caption{
  (a) MR ratios of CCP-CPP-GMR spin-valves shown in
  Fig. \protect\ref{fig3} are plotted against the
  resistance area product (RA).  
  (b) MR ratios shown in Fig. \protect\ref{fig3} are plotted against
  the ratio of the effective resistances $r_{F}/\tilde{r}_{N}$.
  In both panels, circles indicate the maximum value of the MR ratio
  for each value of $\rho_{\text{Cu}}$.
 }
\label{fig4}
\end{figure}

In Fig. \ref{fig4} (a), we plot the MR ratios against the resistance
area product (RA).
This behavior is qualitatively consistent with the experimental
results \cite{fukuzawa04} ,\cite{fukuzawa05}.
The difference of the absolute scale of the MR ratio 
between our theory and experimental results
is due to the absence of high-resistance electrode layers in our
theory, which decreases the MR ratio. 
In the experiment, the total RA of the electrode layers is estimated to be 
about $100 \, \text{m}\Omega \, \mu\text{m}^2$. 
If we take into account this extra resistance, the MR ratio goes down to 10\% 
with the RA of a few hundred $\text{m}\Omega \, \mu\text{m}^2$, 
which quantitatively agrees with the one obtained in the experiment. 

% Moreover it seems that the experimental results are well fitted 
% by our data for $\rho_{\text{Cu}}=17$ and $100 \, \Omega$nm, 
% while in Ref. \cite{fukuzawa05} the experimental data are fitted by
% the theoretical curve for the value $\rho_{\text{Cu}}=1600$ and $650
% \, \Omega$nm.  The discrepancy between expedifference is due to the
% fact that the information of the geometry and 
% resulting effect of bulk spin-dependent scattering is not taken into
% account in Ref. \cite{fukuzawa05}. 

%============================================================
% Fig. 4
%============================================================

The value of the contact radius that maximize the MR ratio can be
qualitatively explained by considering the effective circuit shown in
Fig.\ref{fig1} (b).
For the CCP-CPP-GMR spin-valve, the resistance of the NOL layer
increases with decreasing the radius of the contact. 
In the effective circuit model shown in
Fig.\ref{fig1} (b), the change of the contact radius can be
realized by the corresponding change of effective resistance.
Since the thickness of the N-layer $t_N$ is much  smaller than the
spin diffusion length $\lambda_N$, we can employ the following
approximations: $r_N(P)\sim\tilde{r}_N=\rho_N t_N/(2d^2)$ and $r_N(AP)\sim\infty$, 
where $d$ is the contact radius. 
Based on this consideration, 
in Fig. \ref{fig4} (b), we plot the MR ratio against the ratio of the
effective resistances $r_F/\tilde{r}_N$.
It is observed that 
the MR ratio is maximized when the effective resistances $r_F$ and $\tilde{r}_N$ 
are balanced with each other, i.e.,  $r_F\sim \tilde{r}_N$. 

%============================================================
% Interfacial resistance
%============================================================
Finally we shall discuss the effect of the 
spin-dependent interfacial scattering which is represented by 
the spin-dependent interfacial resistance $r_s$ in the macroscopic
model of the CPP-GMR \cite{valet93}. 
The boundary conditions for electro-chemical potentials at the
interface located at $z=z_0$ are given by
\begin{equation}
\mu_s(z=z_0+0)-\mu_s(z=z_0-0)=r_s\,j_s^z(z=z_0),
\end{equation}
where $j_s^z$ is the $z-$component of the current density and the
spin-dependent interfacial resistance $r_s$ are defined as
\begin{equation}
r^{-1}_{\uparrow}=(1+\gamma)r_b^{-1}/2, \quad r^{-1}_{\downarrow}=(1-\gamma)r_b^{-1}/2, 
\label{ir}
\end{equation}
where $\gamma$ is the interfacial spin asymmetry coefficient and $r_b$
is the total interfacial resistance. 

In our calculation, the spin-dependent interfacial scattering is
realized by introducing fictitious layers with thickness $t_{\text{IR}}$, 
resistivity $\rho_{\text{IR}}=r_b/t_{\text{IR}}$, 
spin polarization of the conductivity $\beta_{\text{IR}}=\gamma$, 
and with the infinite spin diffusion length $\lambda_{\text{IR}}\to\infty$. 

We carried out the same calculations as shown in Figs \ref{fig3} taking
into account of the spin-dependent interfacial scattering at both
sides of the contact.
The parameters are taken to be 
$\rho_{\text{IR}} = 200\,\Omega\text{nm}$ and $\gamma = 0.62$ 
\cite{fukuzawa04}, \cite{fukuzawa05}.
The results are qualitatively same as those shown in in Figs \ref{fig3}.
However, the contact radius that maximize the MR ratio decreases due
to the spin-dependent interfacial scattering.
For $\rho_{\text{Cu}}=17\,\Omega\text{nm}$, for example, the contact
radius that maximize the MR ratio decreases as $3.0$nm $\to 2.3$nm.

%============================================================
% Conclusion
%============================================================
\section{Conclusion}
In conclusion, we theoretically study the magnetoresistance of a
CCP-CPP-GMR system by numerically solving the diffusion equations for
spin-dependent electro-chemical potentials in three-dimensional CCP
geometry.  We show that the MR ratio is not a monotone function of
the contact radius but takes a maximum value at a certain
value of the contact radius.
We also show that the value of contact radius that maximize the MR ratio 
is qualitatively understood by considering matching of the
effective resistances.

%%%%%%%%%%%%%%%%%%%%%%%%%%%%%%%%%%%%%%%%%%%%%%%%%%%%%%%%%%%%%%%%%%%
%%%                       Acknowledgement                       %%%
%%%%%%%%%%%%%%%%%%%%%%%%%%%%%%%%%%%%%%%%%%%%%%%%%%%%%%%%%%%%%%%%%%%
%
\section*{Acknowledgement}
The authors thank M.~Sahashi, M.~Doi, H.~Iwasaki, M.~Takagishi,
Y.~Rikitake and K.~Seki for valuable discussions.  The work
has been supported by The New Energy and Industrial Technology
Development Organization (NEDO).
%%%%%%%%%%%%%%%%%%%%%%%%%%%%%%%%%%%%%%%%%%%%%%%%%%%%%%%%%%%%%%%%%%%
%%%                      Bibliography                           %%%
%%%%%%%%%%%%%%%%%%%%%%%%%%%%%%%%%%%%%%%%%%%%%%%%%%%%%%%%%%%%%%%%%%%
\ifCLASSOPTIONcaptionsoff
  \newpage
\fi

\end{document}